\documentclass[twoside]{dis04}
\def\runauthor{Kamil Sedl\'{a}k}
\def\shorttitle{Dijet Production at Low $Q^2$ at HERA}
\begin{document}

\title{Measurement of Dijet Production at Low
$\mathbf{Q^2}$ at HERA}

\author{Kamil Sedl\'ak}

\address{{\rm (on behalf of the H1 collaboration)}\\
Institute of Physics AS\,CR\\
Na Slovance 2, 182 21 Praha 8, Czech Republic\\
E-mail: ksedlak@fzu.cz}

\maketitle

\abstracts{
A recent H1 measurement~\cite{Aktas:2004px} 
of triple differential dijet cross sections 
in $e^\pm p$ interactions in the region
of photon virtualities $2 < Q^2 < 80\, \mathrm{GeV}^2$ 
is presented and compared to LO and NLO QCD
predictions.  
Effects that go beyond the fixed-order NLO QCD
calculations are identified.
}

\section{Introduction}
Jet cross sections in electron-proton collisions are successfully
described by next-to-leading order (NLO) QCD calculations in most
of the HERA kinematic range.  However, regions of phase space
have previously been observed for which NLO predictions do not
reproduce the data satisfactorily~\cite{Chekanov:2004hz} 
and leading order (LO) Monte
Carlo simulations with different approaches to modelling higher
order QCD effects are often more successful.

In this analysis we focus on the dijet cross section
in a region where we expect large deviations between
measurement and the QCD predictions. 
This kinematic region is characterised by low photon virtualities, $Q^2$, 
forward jet pseudorapidities, $\eta^*$, and high inelasticities,
$y$, (which corresponds to the region of low $x^{\mathrm{jets}}_\gamma$, 
the fraction
of the photon four-momentum carried by the parton involved in 
the hard scattering).

The analysis is described in much more detail in a recent
H1 publication~\cite{Aktas:2004px}.  Here we just recall its most important
results, and in addition present one particular aspect of the analysis, 
namely an unphysical dependence of the NLO QCD predictions calculated using
the program JETVIP~\cite{Jetvip} on a technical parameter $y_c$.
Figure\,\ref{qe3x.ycut}, showing the problematic
$y_c$ dependence, is the only result which is
not included in~\cite{Aktas:2004px}.

\section{Data samples and event selection}
The present analysis is based on a 57\,pb$^{-1}$ data sample taken
in the years 1999 and 2000 with $\sqrt{s}=318\, \mathrm{GeV}$.  
The kinematic region is defined by the cuts: 
$2 < Q^2 < 80\, \mathrm{GeV}^2$; 
$0.1 < y < 0.85$;
$E^*_{T\, 1} > 7\, \mathrm{GeV}$; 
$E^*_{T\, 2} > 5\, \mathrm{GeV}$;
$-2.5 < \eta^*_{1} < 0$;
$-2.5 < \eta^*_{2} < 0$, 
where jet transverse energies, $E^*_T$, and pseudorapidities,
$\eta^*$, are calculated relative to the $\gamma^* p$ collision
axis in the $\gamma^* p$ centre-of-mass frame.  The jets
are ordered according to their transverse energy,
with jet 1 being the highest $E^*_T$ jet.
The dijet cross sections are presented as a function of
the variable $x^{\mathrm{jets}}_\gamma$
defined by eq.\,(1) of~\cite{Lightwood}.

\section{Results}
\begin{figure}\centering
\epsfig{file=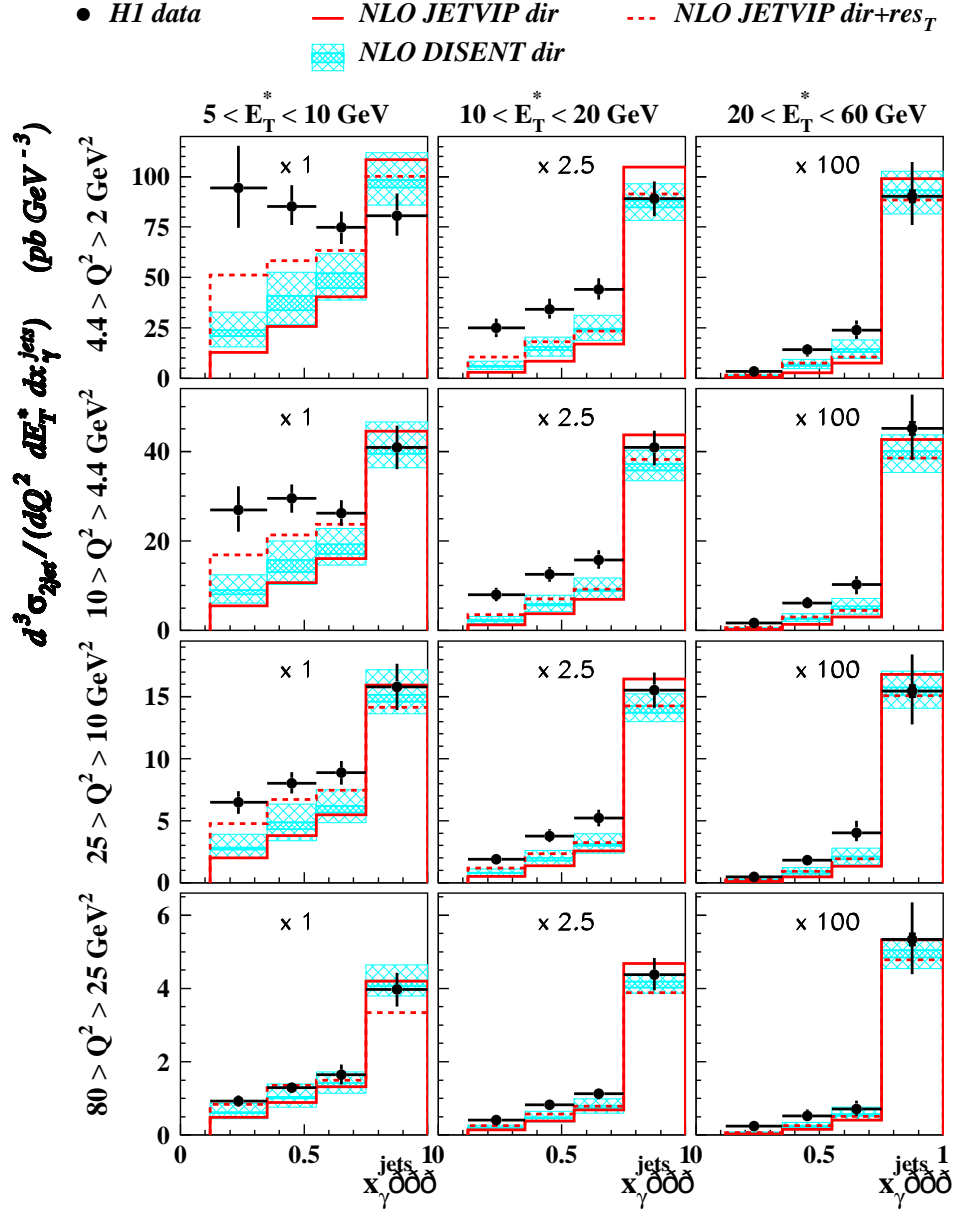,width=\linewidth,%
bbllx=5pt,bblly=15pt,bburx=520pt,bbury=673pt,clip=}
\caption{Triple differential dijet cross section
${\rm d}^3\sigma_{\rm {2jet}}/{\rm d}Q^2 {\rm d}
E^*_T\, {\rm d} x^{\mathrm{jets}}_\gamma$
      with asymmetric $E^*_T$ cuts (see text).
      The inner error bars on the data points show the statistical error,
      the outer error bars show the quadratic sum of systematic
      and statistical errors.
      Also shown are NLO direct photon calculations using DISENT
      (hatched area) and JETVIP (full line),
      as well as the sum of NLO direct and NLO resolved photon contributions
      from JETVIP (dashed line). All calculations are corrected for
      hadronisation effects.
      The inner hatched area illustrates
      the uncertainty due to the hadronisation corrections.
      The outer hatched area shows the quadratic sum of the
      errors from hadronisation and the scale uncertainty (shown
      only for DISENT).
      The scale factors applied to the cross sections are given.}
\label{qe3x.dis}
\end{figure}

Triple differential dijet cross section is presented
as a function of $x^{\mathrm{jets}}_\gamma$ in different bins of $Q^2$ and
$E^*_T$ in Fig.\,\ref{qe3x.dis}.  The variable $E^*_T$ denotes the transverse
energies of the jets with the highest and second highest
$E^*_T$, so that each event contributes twice to the distributions,
not necessarily in the same bin.  The data are compared with
the NLO direct photon calculations performed with DISENT
and JETVIP.  The uncertainties from variations of the 
factorisation and renormalisations 
scales\footnote{$E^*_{T\, 1}$ is taken as the renormalisation
scale, see~\cite{Aktas:2004px} for the factorisation scale.}
in the interval
$\mu/2$ to $2\mu$, as well as from hadronisation corrections,
are illustrated.

Figure\,\ref{qe3x.dis} demonstrates that the NLO direct photon calculations
describe the data in the region of high $x^{\mathrm{jets}}_\gamma$, 
where direct
photon interactions dominate.  For $x^{\mathrm{jets}}_\gamma<0.75$, the
description is nowhere perfect, indicating the need for orders
beyond NLO.  The description of the data for $x^{\mathrm{jets}}_\gamma<0.75$
gets worse as $Q^2$ and $E^*_T$ decrease.  The discrepancy
is particularly pronounced at small $x^{\mathrm{jets}}_\gamma$, 
low $Q^2$ and
low $E^*_T$, where the data lie significantly above the theoretical
predictions, even taking into account the sizable scale
uncertainty.

\begin{figure}\centering
\epsfig{file=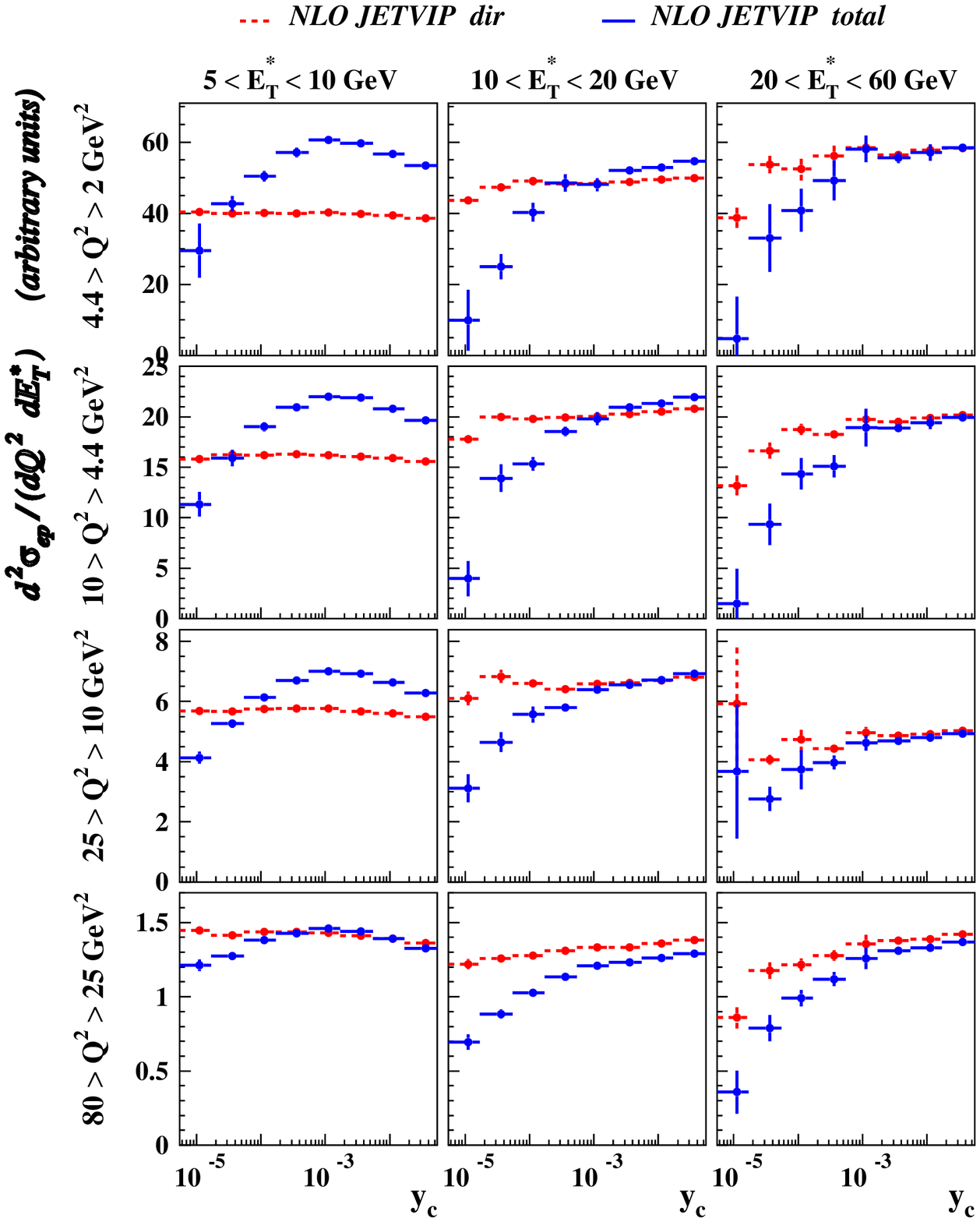,width=\linewidth,%
bbllx=5pt,bblly=15pt,bburx=520pt,bbury=673pt,clip=}
\caption{Dependence of dijet cross section
${\rm d}^2\sigma_{\rm {2jet}}/{\rm d}Q^2 {\rm d} E^*_T$
computed using JETVIP on the parameter $y_c$.
The dotted points show the NLO direct photon contribution,
while the full points are the complete NLO JETVIP prediction 
including the NLO resolved photon contribution.  The former
corresponds to the full points in Fig.\,\ref{qe3x.dis}, the
latter to the dashed points in Fig.\,\ref{qe3x.dis}.}
\label{qe3x.ycut}
\end{figure}

The discrepancy between DISENT and JETVIP direct photon
predictions in Fig.\,\ref{qe3x.dis}
is observed only for multi-differential distributions
which include a jet variable.  It gets substantially
smaller for the inclusive dijet cross section 
${\rm d}^2\sigma_{\rm {ep}}/{\rm d}Q^2 {\rm d}y$~\cite{Thesis} 
and agrees within 2\% for
the total dijet cross section in our kinematic region.

The pattern of the observed discrepancy between the data
and the NLO direct photon calculations in Fig.\,\ref{qe3x.dis} 
suggests an explanation in terms of the interactions of resolved
virtual photons, understood as an approximation to 
contributions beyond NLO.  Of the NLO parton level
calculations, only JETVIP includes a resolved virtual
photon contribution.  Once it is included, the NLO
prediction gets closer to the data, though there is
still a discrepancy between the data and the calculations
at low to moderate $x^{\mathrm{jets}}_\gamma$ and low $Q^2$.

Unfortunately, there is a large dependence of the NLO
resolved photon contribution calculated by JETVIP 
on the slicing parameter\footnote{$y_c$ is just 
a technical parameter used internally in JETVIP
for numerical integration.}   $y_c$.
The resulting JETVIP predictions are therefore
less reliable.  The $y_c$ dependence of the dijet cross section
predicted by JETVIP is shown  in different
bins of $Q^2$ and $E^*_T$ in Fig.\,\ref{qe3x.ycut}. 
For $10^{-5} < y_c < 10^{-3}$, the $y_c$ range 
recommended~\cite{Jetvip} by the authors of JETVIP,
the NLO direct photon predictions are much more stable
than those for the NLO resolved photon contribution.
In the absence of other calculations of this kind, the data
in Fig.\,\ref{qe3x.dis} were compared with the results of the full
JETVIP calculations using $y_c =0.003$. 

Unlike the parton level calculations, DISENT and JETVIP,
LO MC models take into account initial and final state QCD parton
showers.  With the help of the HERWIG MC 
we show in~\cite{Aktas:2004px} that these effects
significantly increase the dijet cross section
in our kinematic region, especially in the region of low
$Q^2$, low $E^*_T$ and low $x^{\mathrm{jets}}_\gamma$.  
We also indicate the effects
of resolved photons with longitudinal polarisation.
Once they are included, the prediction of the MC program HERWIG
provides a better description of the data (see figures and
discussion in~\cite{Aktas:2004px}).  

The MC program CASCADE~\cite{CASCADE} (not shown), which
is based on the CCFM evolution scheme and unintegrated 
proton parton densities, describes the main qualitative trends in the
data except the $Q^2$ dependence.  The overall quantitative 
description of the data is however worse than that by the 
full HERWIG MC.

\section{Conclusions}
The recent H1 measurement of dijet cross section in the low
$Q^2$ region show clear evidence for effects that go beyond
fixed-order NLO QCD calculations.  The improved description
of the data when including 
QCD parton showers and resolved photon contributions
with longitudinal polarisation is illustrated 
in~\cite{Aktas:2004px}.

\end{document}